# Novel Valence Transition in Elemental Metal Europium around 80 GPa

Bijuan Chen,[1,][*] Mingfeng Tian,[2] Jurong Zhang,[3] Bing Li,[1] Yuming Xiao,[4] Paul Chow,[4] Curtis Kenney-Benson,[4] Hongshan Deng,[1] Jianbo Zhang,[1] Raimundas Sereika,[1] Xia Yin,[1] Dong Wang,[1] Xinguo Hong,[1] Changqing Jin,[5] Yan Bi,[1] Hanyu Liu,[6] Haifeng Liu,[2] Jun Li,[7] Ke Jin,[7] Qiang Wu,[7] Jun Chang [8,][†], Yang Ding,[1,][††] and Ho-kwang Mao[1]

[1]*Center for High-Pressure Science and Technology Advanced Research, Beijing, 100094, China*

[2]*Institute of Applied Physics and Computational Mathematics, Beijing, 100088, China*

[3]*Shandong Provincial Engineering and Technical Center of Light Manipulations & Shandong Provincial Key Laboratory of Optics and Photonic Device, School of Physics and Electronics, Shandong Normal University, Jinan 250358, China*

[4]*HPCAT, x-ray Science Division, Argonne National Laboratory, Argonne, IL 60439, USA*

[5]*Institute of Physics, Chinese Academy of Sciences, Beijing 100190, China*

[6]*International Center for Computational Method and Software, College of Physics, Jilin University, Changchun 130012, China*

[7]*National Key Laboratory of Shock Wave and Detonation Physics, Institute of Fluid Physics, CAEP, Mianyang, 621900, China*

[8]*College of Physics and Information Technology, Shaanxi Normal University, Xi'an 710119, China*

[*] bijuan.chen@hpstar.ac.cn

[†] junchang@snnu.edu.cn

[††] yang.ding@hpstar.ac.cn

Valence transition could induce structural, insulator-metal, nonmagnetic-magnetic and superconducting transitions in rare-earth metals and compounds, while the underlying physics remains unclear due to the complex interaction of localized 4*f* electrons as well as their coupling with itinerant electrons. The valence transition in the elemental metal europium (Eu) still has remained as a matter of debate. Using resonant x-ray emission scattering and x-ray diffraction, we pressurize the states of 4*f* electrons in Eu and study its valence and structure transitions up to 160 GPa. We provide compelling evidence for a valence transition around 80 GPa, which coincides with a structural transition from a monoclinic (*C*2/*c*) to an orthorhombic phase (*Pnma*). We show that the valence transition occurs when the pressure-dependent energy gap between 4*f* and 5*d* electrons approaches the Coulomb interaction. Our discovery is critical for understanding the electrodynamics of Eu, including magnetism and high-pressure superconductivity.

Understanding the behaviors of 4*f* electrons is key to elucidating the paradigmatic physical phenomena in lanthanide elemental metals and compounds but remains a long-standing challenge in many-body quantum physics for electron correlated materials [1-7]. The valence transition induced by the changes of external parameters is predominantly associated with the changes of 4*f* electron states [8-10], providing a unique opportunity to investigate the electrodynamics of 4*f* electrons [11, 12].

Among the rare-earth elemental metals, Eu and Yb are distinctive with their divalent state ($Eu^{2+}$-$4f^7$) ($Yb^{2+}$-$4f^{14}$) and larger molar volumes, owing to their half-filled or full-filled 4*f* orbitals [13-15]. Applying sufficient pressure could lead to the delocalization of 4*f* electrons to make Eu and Yb trivalent metals [14, 16-18]. Yb is reported to undergo a continuous evolution from divalent $4f^{14}$ to mixed valence state of $4f^{14}$ and $4f^{13}$ at ~125 GPa [19-22]. In contrast, the valence state of Eu under high pressure is still debated. For instance, Röhler [23] reported the valence of Eu increases from 2 to ~2.5 at around 12 GPa and then becomes saturated (~ 2.64) up to 34 GPa, whereas Bi et al. concluded that Eu retains divalent up to 87 GPa [24-26]. Both Eu and Yb show superconductivity around 80-90 GPa [22, 27]. The origin of the superconductivity in Yb can be attributed to the valence fluctuation-induced magnetic instabilities [22, 28], whereas it remains perplexing to understand how the magnetic collapse and superconductivity could coexist with the strong local spin moments in the divalent Eu metal [25-27, 29]. This problem motivated us to further investigate the valence state of Eu at higher pressure.

In this Letter, we probed the valence transition in Eu using resonant x-ray emission spectroscopy (RXES) up to a record high pressure of ~160 GPa [21, 30-34]. In addition, x-ray powder diffraction (XRD) was also carried out to study if the structural changes are correlated with the valence transitions. As a result, we unveil a novel pressure-induced valence transition in Eu at around 80 GPa, being concomitant with a volume-collapsed structural transition from monoclinic symmetry (*C2/c*) to orthorhombic symmetry (*Pnma*). The valence transition is attributed to the pressure-induced promotion of 4*f* electrons to the 5*d* band, and the valence instability could also explain the origin of the possible superconducting transition occurring around this transition pressure. Details of the experimental settings are provided in the Supplemental Materials (SM) [35].

In our RXES measurements, an electron from the $2p_{3/2}$ core level is photoexcited to an empty $5d_{5/2}$ state ($L_3$ absorption), followed by the decay of an electron from $3d_{5/2}$ state to fill the $2p_{3/2}$ core hole ($L_\alpha$ emission). According to the Anderson impurity model [55-58] the cross section of this two-step core-core resonant inelastic scattering is proportional to the unoccupied density of 5*d* states that is convoluted with a many-body expectation value including 2*p* and 3*d* core holes. Even though the 4*f* states are not directly involved in the excitations, the core hole in the 3*d* state modifies the total energy of the localized 4*f* electrons [57, 58]. When more than one $4f^n$ configuration is mixed in the initial state, the modification splits $4f^n$ configurations in the absorption edge to yield valence histogram information [57, 58].

Figure 1 depicts the RXES measured on Eu at 11 GPa as a function of the energy transfer $E_t$ (defined as incident energy $E_i$ – outgoing energy $E_o$), as well as a partial yield fluoresce x-ray absorption spectrum (PYF XAS) collected in the absorption mode with the $E_o$ fixed at 5846 eV. Due to the 3*d* core-hole effects, two peaks are identified in RXES at around 1128 and 1135 eV with an energy separation of ~ 7 eV, which are associated with the final states $3d^9 4f^7 5d^1$ (labeled as $4f^7$) and $3d^9 4f^6 5d^1$ (labeled as $4f^6$) [59], respectively. As peak $4f^6$ shows a more prominent line-shape at $E_i$ = 6970 eV, we use this RXES spectrum to monitor how the valence of Eu evolves at high pressure in this study.

Figure 2(a) shows ten RXES spectra collected with $E_i$ = 6970 eV from 11 to 160 GPa. The spectra are normalized to the peak $4f^7$ maximum intensity. Analysis with the Gaussian peak fitting yields the intensity of peaks $4f^7$ and $4f^6$. The valence is estimated using the conventional formula (1) for RXES, and XANES [23] measurements,

$$v = 2 + \frac{I(4f^6)}{I(4f^6)+I(4f^7)} \quad (1)$$

where $I(4f^7)$ and $I(4f^6)$ are the area integrated intensities of $4f^7$ and $4f^6$ peaks, respectively [21, 60, 61]. Figure 2(b) shows the resulting valence state. The errors are primarily due to statistics of total counts and fitting errors, which are estimated to be within ~5%. It is worth noting that a valence jump appears around 80 GPa and then gradually increases up to 160 GPa, indicating that a valence transition begins around 80 GPa.

In addition, we performed XRD up to 153 GPa to investigate the structural changes. The results (Fig. 3) show that Eu experiences a phase transition from a body-center-cubic (bcc) to a hexagonal-closed-packing (hcp) structure at ~ 12 GPa, with a ~ 3% volume collapse, in agreement with previous studies [24, 62-64]. Eu remains stable in the hcp phase from 12 GPa up to 30.1 GPa and then transforms into an incommensurately modulated monoclinic crystal structure with symmetry of *C2/c*, as reported by Husband et al. [64]. When pressure exceeds 78 GPa, a new reversible structural phase transition with a 3.2% volume collapse occurs. The new phase is stabilized in an orthorhombic crystal structure with symmetry of *Pnma*, according to the structural refinement of the XRD pattern at 96 GPa (Fig. 3(b)). The bulk modulus ($B_0$) and pressure derivative of the bulk modulus ($B_0$') are determined as 13.25 GPa and 2.29 (see Fig. S6). These low values of $B_0$ and $B_0$' are comparable with those observed in Yb [65]. This unusually high compressibility of Eu is possibly associated with the valence transition [65].

So far, the valence state of Eu below 80 GPa has remained a point of contention. Röhler et al. discovered that pressure significantly suppresses the $4f^7$ peak while only slightly increasing the $4f^6$ peak in their XANES experiments up to 34 GPa [23]. Using the formula (1), they determined that the valence changes from 2 to 2.64 around 34 GPa, despite of the fact that the change is primarily due to the suppression of the $4f^7$ peak. Bi et al., on the other hand, attribute the changes in the $4f^7$ peak below 87 GPa to pure $5d$ states and conclude Eu retains in a nearly divalent state up to 87 GPa [25]. We confirmed these findings with PYF XAS measurements (Fig. S4): the $4f^7$ peak is entirely suppressed below 52 GPa, while the $4f^6$ peak grows very slightly below 80 GPa.

In the previous studies of lanthanide compounds [66-69], the decrease of the $4f^7$ peak commonly results in a corresponding increase of the $4f^6$ peak, and the total weight of the $4f^7$ peak and $4f^6$ peak in the transitions remains approximately constant [30, 70]. This association between the $4f^7$ and $4f^6$ peaks validates the use of formula (1) to estimate the valence state. Because no such connection exists in the XANES and PYS-XAS of Eu at high pressures up to 147 GPa, it implies that the XANES and PYF XAS may not be good probes for studying the valence in Eu owing to several difficulties stated below.

The intensity of the $L_3$-edge white line, which overlaps with the $4f^7$ state in the XANES measurements, is dominated by the density of $5d$ states (due to the selection rule) and thus strongly influenced by the change of $5d$ states rather than $4f^7$ state. Furthermore, the intensity of the white line is sensitive to changes in sample thickness, defects, inhomogeneity, as well as pressure gradient. As a result, the change in intensity of the white line alone is insufficient to evince a valence transition. In addition, the step-function-like background above the absorption edge and a strong fluorescence background in the PYF XAS above absorption edge may cause uncertainties in resolving the $4f^6$ peaks.

In contrast, the RXES spectra measured below the absorption edge avoid the problems arising from white line and fluorescence background and are regarded as a superior probe for studying the valence transition of rare-earth metals and compounds at high pressure [21, 60, 61, 71]. In our RXES measurements, the sum of $I(4f^7)$ and $I(4f^6)$ remain nearly a constant at high pressure up to 160 GPa, showing a clear correlation between $4f^7$ and $4f^6$ peaks (Fig. S3). Therefore, the significant increase ~35 % (from 2.2 to 2.4 relative to the total valence increase) of valence state around 80 GPa provides conclusive evidence of a valence transition in Eu. In contrast, from 11 GPa to 52 GPa, the valence increases only by ~1 % (from 2.19 to 2.21), showing no evidence of a valence transition.

Consequently, Eu's phase space can be divided into four zones (from I to IV) based on its crystal structures and valence states, as shown in Fig. 2(b). Even though Eu experiences three structural changes and one magnetic transition [26], the $f$ electrons stay nearly localized below the 80 GPa areas (from the region I to III). From 80 to 160 GPa, Eu changes into an orthorhombic structure, with the valence fast increasing to 2.4 about 80 GPa and then gradually increasing to 2.56 around 160 GPa. It is worth noting that the magnetic ordering collapses about 80 GPa [26], while the possible superconducting transition is reported to occur around 75 GPa [27]. Considering the 10 % uncertainties in pressure calibrations from separate studies, the valence transition, magnetic transition, and the possible superconducting transition [27] are likely to coexist. Above 80 GPa, Eu remains in a mixed-valence state. Assuming that the valence increases asymptotically above 160 GPa in the same way as for other $4f$ materials [72], it is extrapolated to reach trivalency near 380 GPa.

So far, three theoretical models have been proposed to account for the mechanisms of the valence transitions. Namely, (i) the *promotional model*, in which the $4f$ electron jumps into the $5d$- electron conduction band to induce a valence transition $4f^75d^0 \rightarrow 4f^65d^1$ [73, 74], (ii) the *Mott-transition model* where the Mott-Hubbard gap is closed and 4f electrons become itinerate coherently among all lattice sites forming a valence fluctuation $4f^7 \rightarrow 4f^6$ [75], and finally (iii) the *Kondo model* where the 4f electrons couples with *spd*-conduction electrons to form Kondo singlets either at a single site or coherently at all sites (Kondo lattice) [76]. However, as no Kondo effect is observed in Eu and the local magnetic moment remains nearly the same [26], the Mott transition model is also unsuitable for explaining the valence transition. Considering 5d state is dominant at the Fermi level [17], and the 4f state locates about 2 eV below the Fermi level at ambient pressure [77], it is likely that the 4f state approaches 5d states and induces the valence transition at ~ 80 GPa, fitting into the promotional model.

If we only consider the conducting 5d orbital and the localized 4f orbital bands in the valence transition, we can understand the valence transition using a Hund-Heisenberg-like model [78],

$$H = H_d + H_f - J_h \sum_i \mathbf{S}_{di} \cdot \mathbf{S}_{fi} + J_H \sum_{\langle i,j \rangle} \mathbf{S}_{fi} \cdot \mathbf{S}_{fj} \qquad (2)$$

where the effective spin operators are $\mathbf{S}_{di} = d_i^\dagger \boldsymbol{\sigma} d_i / 2$ and $\mathbf{S}_{fi} = f_i^\dagger \boldsymbol{\sigma} f_i / 2$ with the Pauli vector $\boldsymbol{\sigma}$. $J_h$ is the Hund coupling between the 5d electrons and the localized 4f electrons at the same site [29], and $J_H$ is the Heisenberg interaction between the *f* orbital electrons on the Eu lattice. The carrier energy in each site of the lattice $\varepsilon_d = \varepsilon_{d0} + \langle n_{fi} \rangle U_{df}$ in $H_d$ includes the energy renormalization from the Coulomb interaction between the *d* and *f* electrons under the mean field approximation. Similarly, $\varepsilon_f = \varepsilon_{f0} + \langle n_{di} \rangle U_{df}$. Once the $4f^7 \rightarrow 4f^6$ valence transition occurs, the $\varepsilon_d$ shifts to lower energy and $\varepsilon_f$ is elevated.

The divalent Eu with $4f^7$ electron configuration possesses a strong local magnetic moment with $J$ = 7/2, and the trivalent-$4f^6$ state is nonmagnetic or $\langle \mathbf{S}_{fi} \rangle = 0$ since $J = L - S = 0$ with $S = L = 3$.

When pressure increases, the hopping integrals normally increase, and so do the widths of $d$ bands and $J_H$, while $J_h$ is usually insensitive to the pressure. Before the valence transition, there is no contribution from the Hund's interaction as there is no electron on the 5$d$ orbitals. After the valence transition, there is no contribution from Hund's interaction either as the 4$f^6$ state has no spin moment.

Figure 4 illustrates the schematic of the promotional model in which the valence transition is associated with the charge transfer between 4$f$ and 5$d$ states. We define an energy gap, $\Delta_{f7} = \varepsilon_d - \frac{D_d}{2} - \varepsilon_f$, between the $d$ and $f$ band for the 4$f^7$ configuration. The contribution from the Heisenberg interaction is much weaker than other terms; thus, this term can be ignored in our following energy calculations. The onsite 4$f$-5$d$ charge transfer induces the local energy change $E(4f^6 5d^1) - E(4f^7 5d^0) \sim \Delta_{f7} - U_{df}$, while the intersite 4$f$-5$d$ charge transfer changes the energy $\Delta_{f7} - J_h \langle S_{di} \cdot S_{fi} \rangle$ (more details can be found in the SM). When the pressure exceeds the critical value of valence transition as $\Delta_{f7} - U_{df} \leq 0$, the valence-transition related phase transition strongly suppresses both the Hund coupling and Heisenberg coupling, giving rise to a metal-like system with $\varepsilon_d - \frac{D_d}{2} < \varepsilon_f$. According to RXES, as the 4$f$ level increases about 0.4 eV from ambient pressure to 80 GPa (see Table S1), and thus $\Delta_{f7}$ decreases to 1.60 eV. By taking $\Delta_{f7} \cong U_{df}$, we obtained valence is about 2.45 using the equation (2) in reference [73], which is close to our measured value 2.4. The possible superconductivity in Eu metal is likely to originate from the valence instability around 80 GPa, and the low $T_c$ value is likely due to the Eu metal not being fully trivalent [27]. Recently, it has been pointed out that the $U_{fd}$ may drive the quantum criticality observed in other strongly correlated electron systems such as Ce and Yb compounds [79-81]. Thus our work provides the important information of the physics underlying the unconventional superconductivity in the strongly correlated electron systems.

In summary, using RXES, we have studied the valence transition in Eu as a function of pressure up to 160 GPa and discovered a new valence transition occurred at around 80 GPa, which is nearly committed with the phase transition from *C*2/c to *Pnma* and the superconductivity transition. The valence transition is driven by the promotion of 4*f* electrons to 5*d* bands. We gave the transition value of the pressure-dependent energy gap between 4*f* and 5*d* electrons which is close to their Coulomb interaction.

Y.D acknowledges the support from the National Natural Science Foundation of China (NSFC) grant No. 11874075, Science Challenge Project No. TZ2016001, and U1930401, and National Key Research and Development Program of China 2018YFA0305703. The RXES, PFY-XAS and XRD experiments were performed at HPCAT (Sector 16), Advanced Photon Source (APS), Argonne National Laboratory. HPCAT operations are supported by DOE-NNSA's Office of Experimental Sciences. The Advanced Photon Source is a U.S. Department of Energy (DOE) Office of Science User Facility operated for the DOE Office of Science by Argonne National Laboratory under Contract No. DE-AC02-06CH11357. H. -K. Mao acknowledges supports from the National Natural Science Foundation of China Grant No. U1530402 and U1930401.

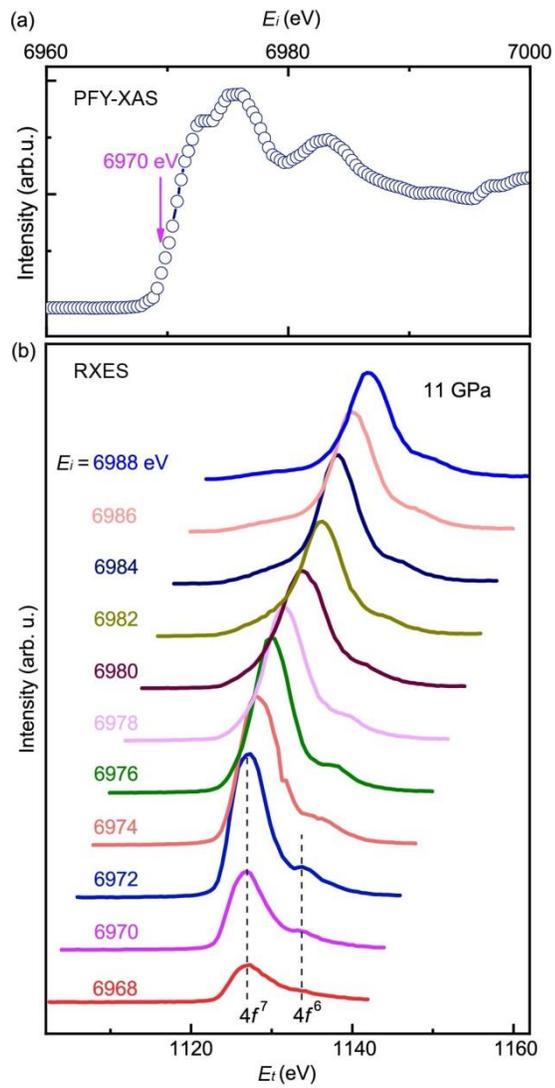

Fig. 1. (a) The normalized $L_3$-edge PFY-XAS spectrum of Eu at 11 GPa. (b) RXES spectra collected at 11 GPa as a function of transfer energy $E_t$ and incident energy $E_i$.

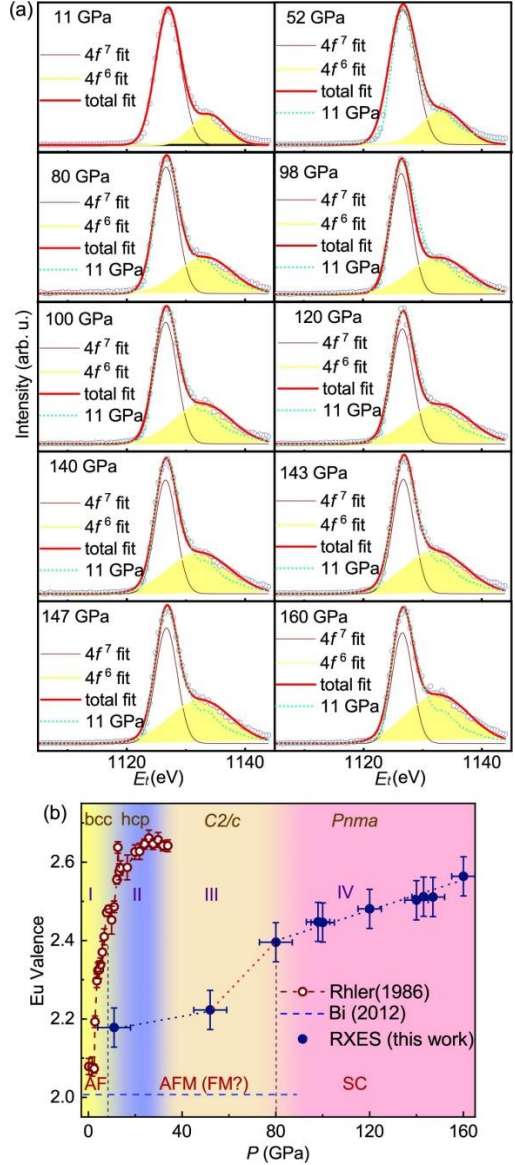

Fig. 2. (a) RXES spectra measured with $E_i = 6970$ eV, which are normalized to the maximum intensity of $4f^7$ peak. (b) Pressure dependence of Eu valence at room temperature as determined by RXES (dark blue solid circles) from this study, and XANES (blue hollow circles) from Ref. [23] and dotted line from Ref. [25], along with the magnetic ground states of Eu from Refs. [25, 27]. The different colors represent different structures of bcc, hcp, monoclinic and orthorhombic as determined by XRD in this study, respectively, which will discuss later. The dashed line is a guide for the eye, where the increase valence around 80 GPa is highlighted by the red zone.

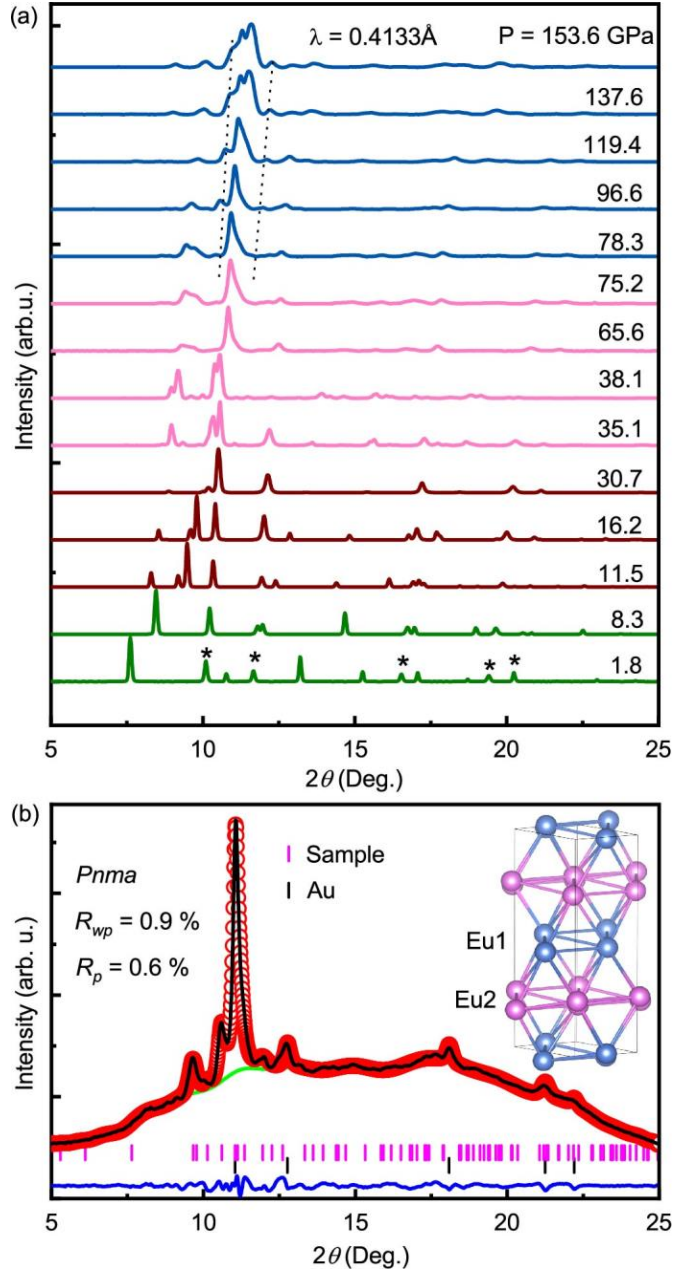

**Fig. 3**. (a) Selected synchrotron XRD pattern with the subtracted background of Eu at various pressures. (b) Rietveld refinement of the XRD patterns collected at 96.6 GPa for the *Pnma* structure at room temperature. The red circles, the solid black line and the green line represent the experimental data, fitted data, and background, respectively. The inset schematic figure shows the local coordination in Eu.

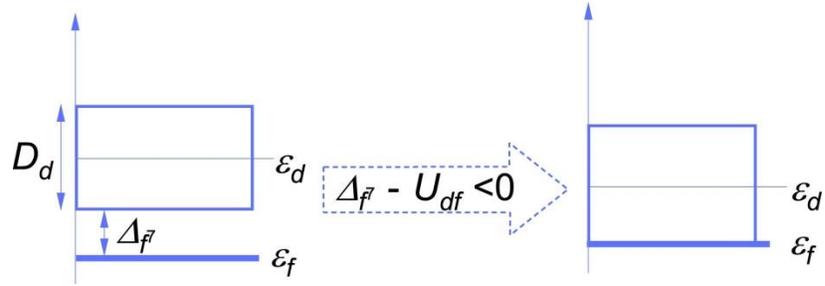

**Fig. 4.** Schematic of the promotional model in which valence transition is associated with the onsite charge transfer between 4*f* and 5*d* states, where $U_{df}$ is the inter-orbital Coulomb repulsion between the carrier and localized hole on the same site, $D_d$ is the bandwidth of 5*d* band. Once the gap approaches $U_{df}$, the valence transition occurs.